\documentclass[twoside]{article}
\usepackage{epsf, latexsym, ijmpa1}

\catcode`\@=11
\long\def\@makefntext#1{
\protect\noindent \hbox to 3.2pt {\hskip-.9pt  
$^{{\eightrm\@thefnmark}}$\hfil}#1\hfill}               

\def\thefootnote{\fnsymbol{footnote}}
\def\@makefnmark{\hbox to 0pt{$^{\@thefnmark}$\hss}}    
        
\def\ps@myheadings{\let\@mkboth\@gobbletwo
\def\@oddhead{\hbox{}
\rightmark\hfil\eightrm\thepage}   
\def\@oddfoot{}\def\@evenhead{\eightrm\thepage\hfil
\leftmark\hbox{}}\def\@evenfoot{}
\def\sectionmark##1{}\def\subsectionmark##1{}}

\newcommand{\disable}[1]{}

\newcommand{\journal}[5]{#1;\textsl{ #2} \textbf{#3 }(#4) #5}
\newcommand{\ba}{\begin{myeqnarray}}
\newcommand{\ea}{\end{myeqnarray}}
\newcommand{\be}{\begin{equation}}
\newcommand{\ee}{\end{equation}}

\newcommand{\oa}{\mathcal{O}(a)}

\newcommand{\bbbar}{b\bar{b}}

\newcommand{\Lower}[2]{\smash{\lower #1 \hbox{#2}}}

\newcommand{\axdi}{$\diamond$}
\newcommand{\axoc}{\Lower{-0.2ex}{$\scriptstyle\bigcirc$}}

\newcommand{\foursquarepictures}[4]{
\begin{picture}(7, 3.7)
  \put(-.05, 1.4){\epsfxsize=2.35in \epsfbox[10 30 560 590]{./#1}}
  \put(2.6, 1.4){\epsfxsize=2.35in \epsfbox[10 30 560 590] {./#2}} 
  \put(-.05, -.6){\epsfxsize=2.35in \epsfbox[10 30 560 590] {./#3}}
  \put(2.6, -.6){\epsfxsize=2.35in \epsfbox[10 30 560 590] {./#4}} 
\end{picture}
}

\newcommand{\twoflatpictures}[2]{
\begin{picture}(7, 1.6)
  \put(-.03, -.75){\epsfxsize=2.35in \epsfbox[10 30 560 590] {./#1}}
  \put(2.58, -.75){\epsfxsize=2.35in \epsfbox[10 30 560 590] {./#2}} 
\end{picture}
}

\def\qed{\hbox{${\vcenter{\vbox{                        
   \hrule height 0.4pt\hbox{\vrule width 0.4pt height 6pt
   \kern5pt\vrule width 0.4pt}\hrule height 0.4pt}}}$}}

\renewcommand{\thefootnote}{\fnsymbol{footnote}}        

\begin{document}

\runninghead
{An investigation into the Fermilab approach to heavy quarks on the lattice.}
{An investigation into the Fermilab approach to heavy quarks on the lattice.}

\normalsize\textlineskip
\thispagestyle{empty}
\setcounter{page}{1}

\copyrightheading{}                     

\vspace*{0.88truein}

\fpage{1}
\centerline{\bf AN INVESTIGATION INTO THE FERMILAB APPROACH}
\vspace*{0.035truein}
\centerline{\bf TO HEAVY QUARKS ON THE LATTICE}
\vspace*{0.37truein}
\centerline{\footnotesize ZBIGNIEW SROCZYNSKI
\footnote{Current address: Theoretical Physics, University of
  Wuppertal, 42097 Wuppertal, Germany.}}
\vspace*{0.015truein}
\centerline{\footnotesize\it 
Department of Physics, University of Illinois, 1110 West
Green Street,}
\baselineskip=10pt
\centerline{\footnotesize\it Urbana, IL 61801, USA.}
\vspace*{10pt}
\vspace*{0.225truein}
\publisher{}{}

\vspace*{0.21truein}
\abstracts{We use a space-time asymmetric $\oa$ improved fermion
  action  and fix the asymmetry non-perturbatively to restore
  the relativistic dispersion relation.
We compute
 spectra and matrix elements of quarkonia and heavy-light mesons
 and compare with results obtained using a symmetric
 action with the Fermilab interpretation \textit{i.e.} that the
 physics of heavy lattice quarks 
depends solely on their kinetic mass. We 
 provide additional evidence to support this.}{}{}


\vspace*{1pt}\textlineskip      
\section{The Fermilab Fermion Action}    
\vspace*{-0.5pt}
\noindent
It is known that the use of a space-time symmetric fermionic lattice action
gives rise to artifacts of order $am_q$ which become significant at
currently accessible 
lattice spacings $a$ and for quark masses $m_q$ large enough for heavy
flavour physics. This becomes manifest in the dispersion relation
for lattice quarks, where a mass can be defined in two different
ways; as the rest mass or the kinetic mass (denoted $M_1$ and $M_2$
respectively). On the relativistic
mass shell these two definitions are equivalent but these lattice
artifacts make them unequal for $am_q>0$.

While there exist a number of approaches to the problems of simulating
heavy quarks on the lattice,\cite{hqmethods}
the Fermilab action arises when the Symanzik improvement
program is applied to the Wilson action in a mass dependent
way.\cite{fnalaction}
This requires that the operators in the action be allowed to be asymmetric
in space and time as well as the full $am_q$  dependence of their
coefficients. 

However, by interpreting a symmetric Wilson action in the light of the Heavy
Quark Effective Theory (HQET) and normalising its operators accordingly
one should be able to do without this asymmetry and use the
standard Sheikholeslami-Wohlert (SW) action). This interpretation
posits that of the two possible lattice quark mass terms, it is only
the kinetic mass which is physically significant and that therefore
this  should be used to set the physical scale.
Hitherto this has been the method used
by the Fermilab and some other groups.
Here, we investigate these claims in further detail.

\pagebreak

\textheight=7.8truein
\setcounter{footnote}{0}
\renewcommand{\thefootnote}{\alph{footnote}}

\section{Comparison of Spectrum and Matrix Elements}
\vspace*{-0.5pt}
\noindent
We implement the asymmetric action and adjust the
asymmetry non-perturbatively until 
$M_1\!=\!M_2$ (within statistical errors). This has been done on
a $12^3\times 
24$ quenched lattice at $\beta=5.7$ and initial results were reported
at \textsl{Lattice99}.\cite{me} Since then, the non-perturbative
fixing of the asymmetry has been
extended from the quarkonium sector to the heavy-light sector and the
statistical errors have been reduced.\footnote{details in a forthcoming
  publication} 
Then we use this action to compute quarkonium and heavy-light mesons
with heavy quark masses corresponding to the $b$ and $c$ quark, and a
light quark at the $s$ mass. We can compare \disable{these spectra and decay
constants} our physical quantities to those obtained from previous
computations on
this same lattice using the SW action where $M_1$ is unconstrained.\cite{fnalresults} 
\begin{figure}[h]
\twoflatpictures{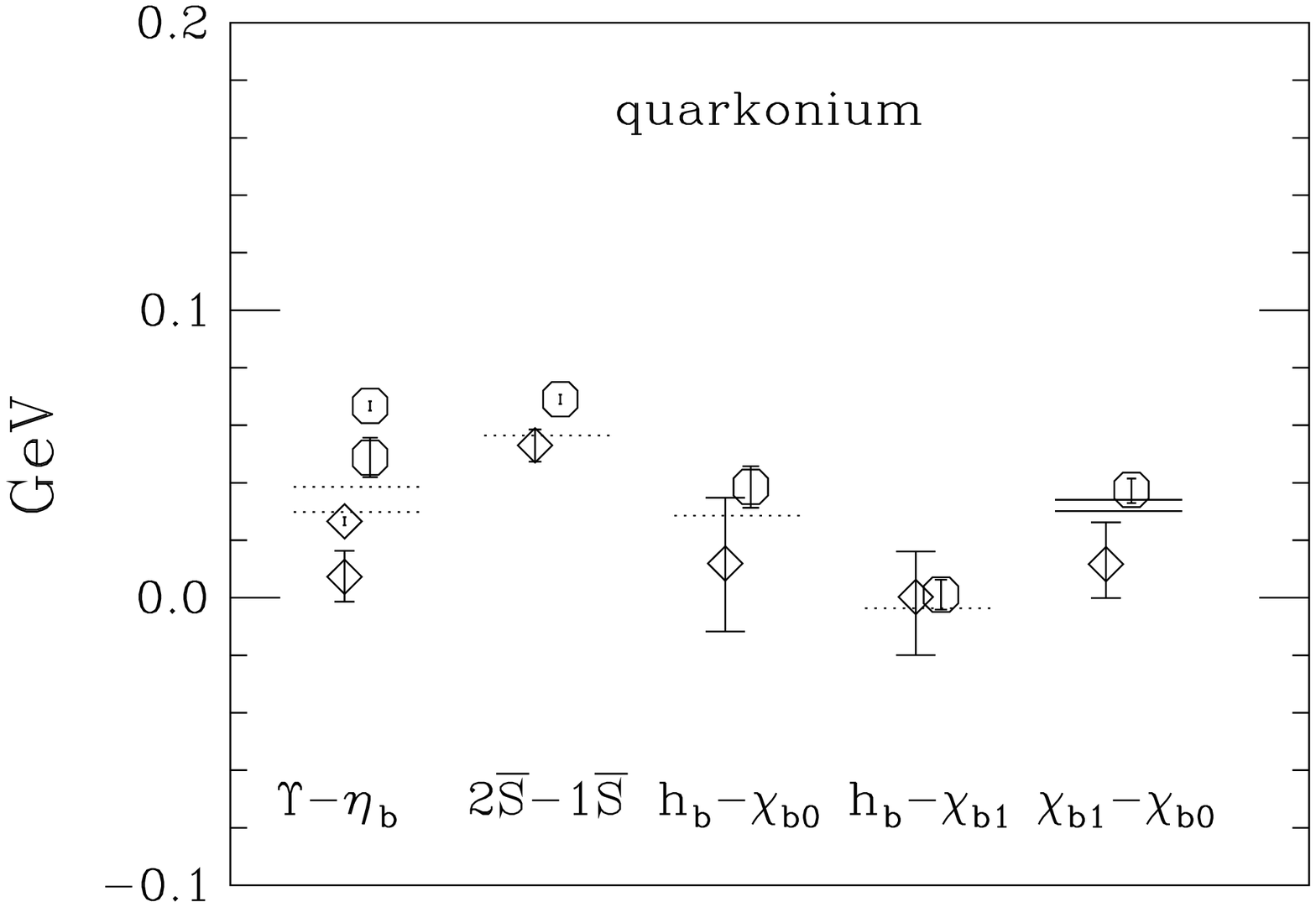}{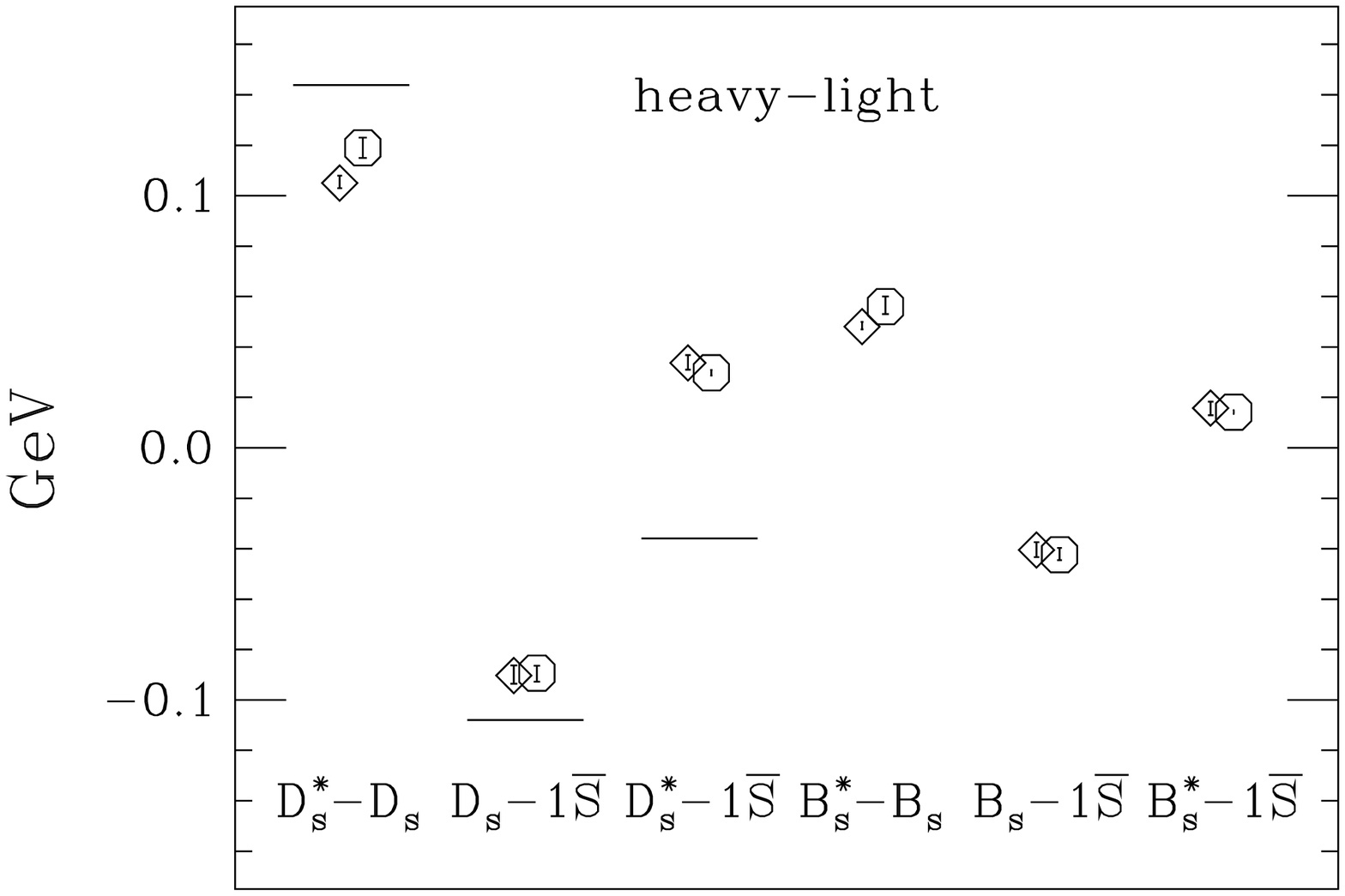}
\caption{\label{spectrum} 
Comparison of results from the asymmetric (\axoc) and the SW (\axdi)
actions together with experimentally measured values.}
\end{figure}

As an example, Fig. \ref{spectrum} shows 
hyperfine, radial and fine structure splittings in the $\bbbar$
system, and, in the heavy-light sector, the hyperfine splittings
and splittings from the spin-averaged $1\bar{S}$ state of the
$c\bar{s}$ system.
Physical quantities from the two actions are in good agreement despite
difference in the values of
$M_1$, and we obtain
a similarly good agreement in the charmonium system and for heavy-light
pseudoscalar current matrix elements,

\section{Kinetic Mass Dependence}
\begin{figure}[ht]
\foursquarepictures{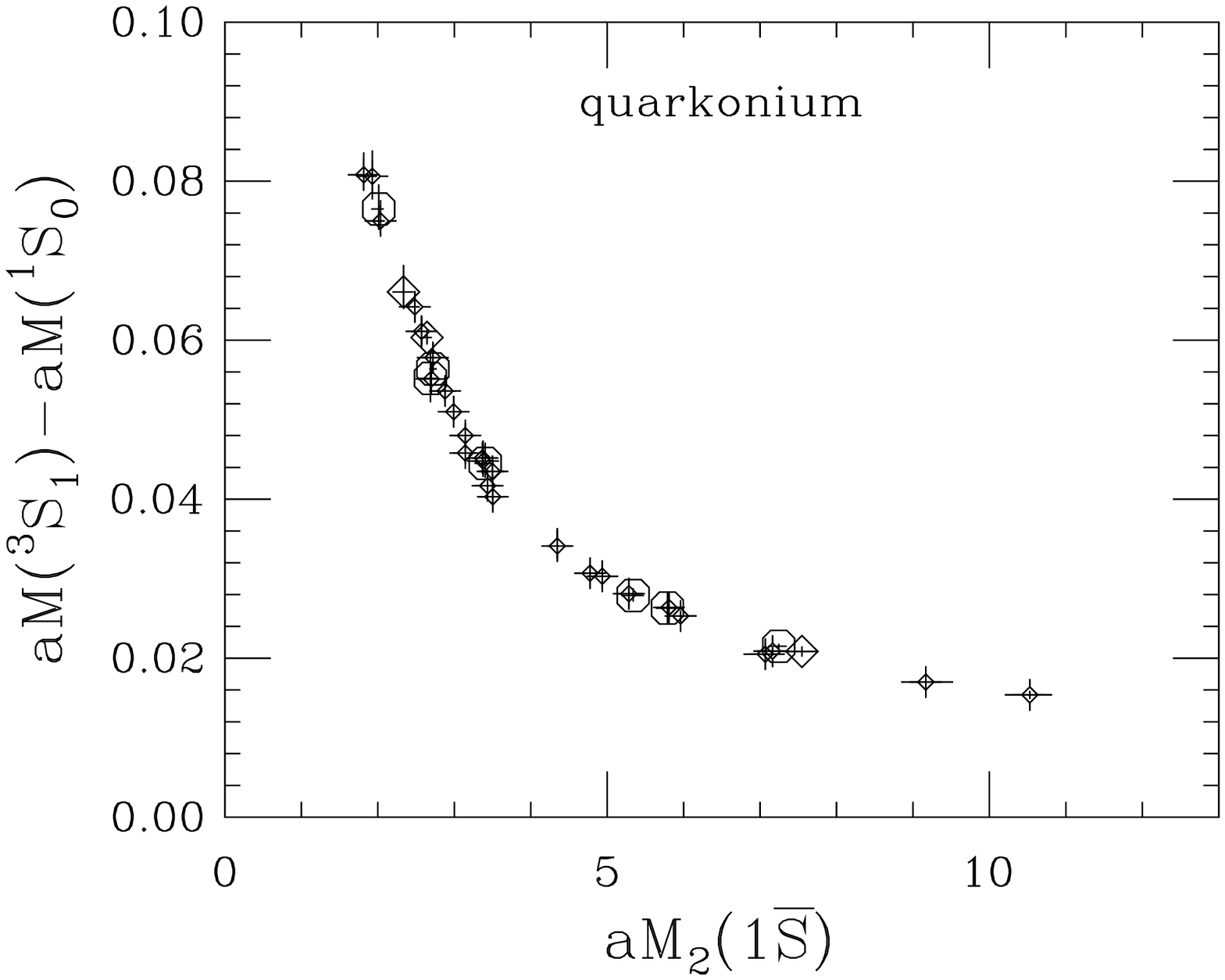}{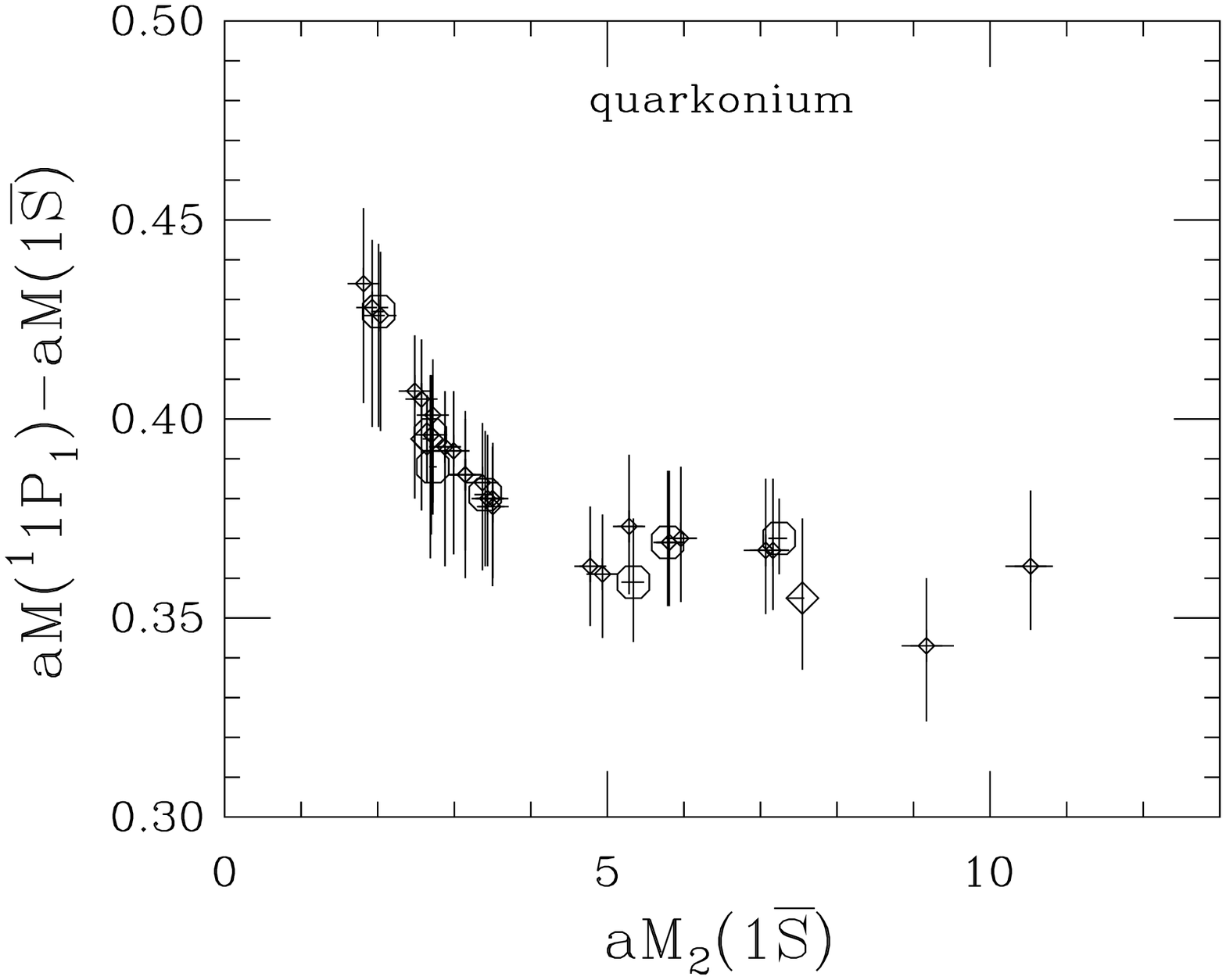}{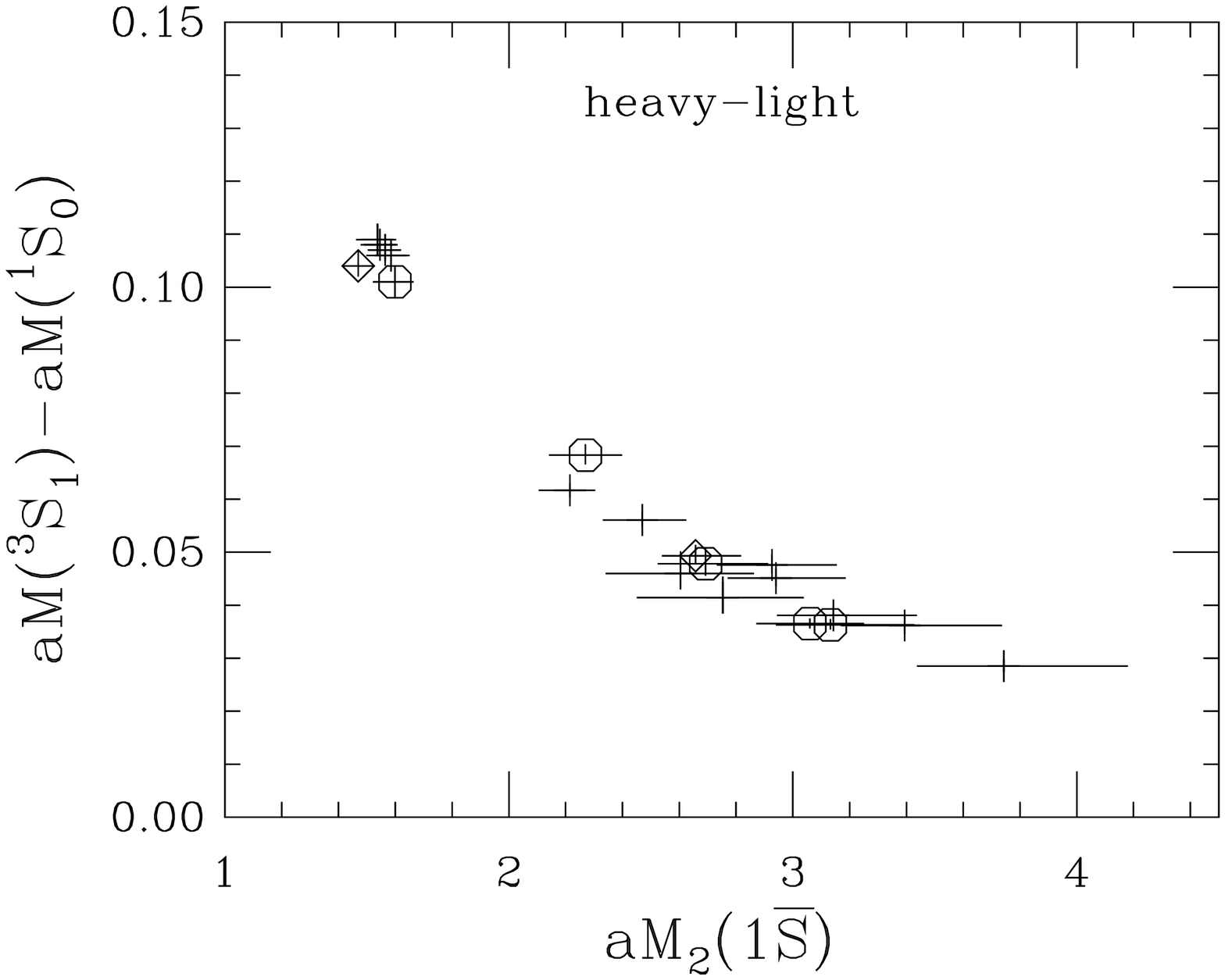}{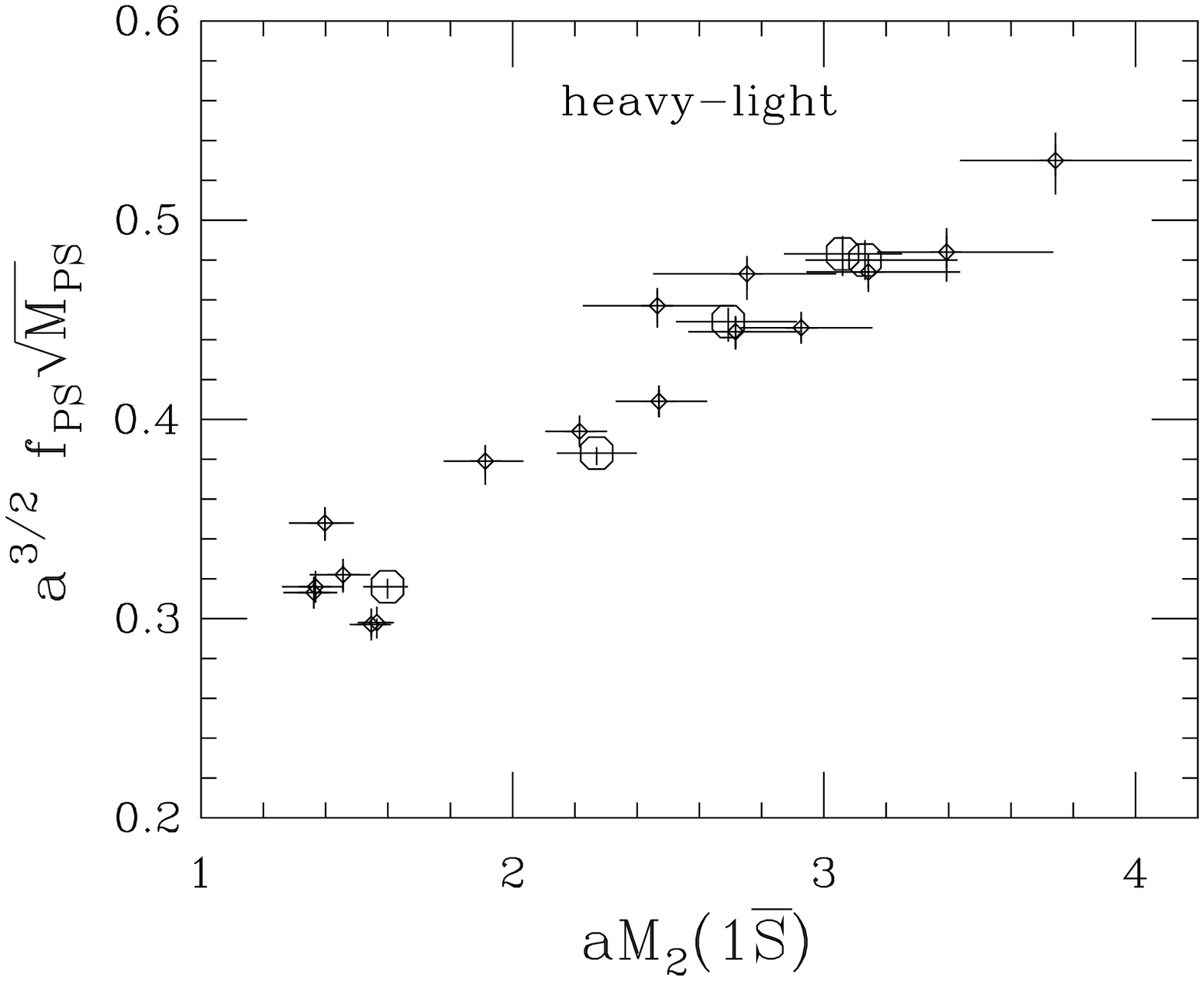}
\caption{\label{m2} 
The dependence on $M_2$ of the quarkonium hyperfine and
 $1P\!-\!1S$ splittings and the heavy-light hyperfine splitting
and the decay constant, using values obtained by using the
 non-perturbatively fixed
(\axoc), the SW (\axdi) and an arbitrarily  asymmetric (+)
action.}
\end{figure}

\vspace*{-0.5pt}
\noindent
The asymmetry of our action allows us to arbitrarily set the ratio of
$M_1$ and $M_2$, and this gives us a useful tool with
which to probe their roles. We compute various mesonic physical quantities 
over a range of kinetic masses and either fix $M_1\!=\!M_2$ or leave
it arbitrary. We add to these results those obtained using the SW
action. As shown in figure \ref{m2} we find that there is clearly a smooth
dependence of these various quantities on $M_2$.

\section{Summary}
\vspace*{-0.5pt}
\noindent
Our results demonstrate that physical quantities depend on $M_2$
rather than $M_1$ and that
the Fermilab prescription for
using the SW action to compute heavy quark physics does indeed bypass the
$\mathcal{O}(am_q)$ artifacts that would otherwise arise.

\nonumsection{Acknowledgements}
\vspace*{-0.5pt}
\noindent
I am grateful for the collaboration on this work of Aida El-Khadra,
Andreas Kronfeld, Paul Mackenzie and Jim Simone.

\nonumsection{References}

\end{document}